\documentclass{article}
\usepackage[latin1]{inputenc}
\usepackage{amsmath,amssymb}
\usepackage{graphicx}
\usepackage{url}
\usepackage{booktabs}

\newcommand{\von}[1]{\\ \footnote{#1}}

\def\XXX#1{}

\def\FUTUREWORK#1{}

\def\Em#1{{\em{#1}\/}}

\def\avg{{\text{\bf avg}}}

\title{Emerging Markets for RFID Traces}
\author{
        Matthias Bauer {\small\tt (matthiasb@acm.org)}
        \von{Shoestring Foundation, Erlangen} \\ \and
        Benjamin Fabian {\small\tt (bfabian@wiwi.hu-berlin.de)} \and
        Matthias Fischmann {\small\tt (fis@wiwi.hu-berlin.de)} \and
        Seda Gürses {\small\tt (seda@wiwi.hu-berlin.de)}
        \von{Institute of Information Systems, Humboldt-University, Berlin} \\ \and
       }

\date{}

\begin{document}

\maketitle
\XXX{\input{tag}}

\begin{abstract}
RFID tags are held to become ubiquitous in logistics in the
near future, and item-level tagging will pave the way for
Ubiquitous Computing, for example  in application fields like smart homes.
Our paper addresses the value and the production cost of 
information that can be gathered by observing these tags over time and different locations.
We argue that RFID technology will induce a thriving market for such
information, resulting in easy data access for analysts to infer
business intelligence and individual profiles of unusually high detail.
%We present a simple model that describes the relationship between the
%various parties on this market and list some basic economic
%properties.

Understanding these information markets is important for many reasons:
They represent new business opportunities, and market players need to
be aware of their roles in these markets. Policy makers need to confirm that the market structure will not negatively affect overall welfare.
Finally, though we are not addressing the complex issue of privacy, we are
convinced that market forces will have a significant impact on the
effectiveness of deployed security enhancements to RFID technology.

In this paper we take a few first steps into a relatively new field of
economic research and  conclude with a list of research problems that promise deeper insights
into the matter.
\end{abstract}

%\XXX{fis@all: abstract has grown a little; i intentionally didn't
%mention EPC or the global uniqueness of traces.  i think these are
%technical details that belong to the introduction.  NEEDS ANOTHER GO
%BEFORE SUBMISSION!!!}

%%%%%%%%%%%%%%%%%%%%%%%%%%%%%%%%%%%%%%%%%%%%%%%%%%%%%%%%%%%%%%%%%%%%%%%%%%%%%%%%%%%%%%%%%%%%%%%%%%%%
\section{Introduction}

RFID tags are widely held to become ubiquitous in our future.  
They are already essential in
logistics \cite{Garfinkel:2005}, and item-level tagging will pave the
way for Ubiquitous Computing applications such as so-called smart
homes \cite{Helal:2005a}.
Our paper is based on a still hypothetical (but widely accepted) Ubiquitous
Computing scenario where goods are tagged on a common basis, and aims
to demonstrate that the incentives in this scenario will induce
thriving markets for information of the form ``serial number $i$ was
at location $l$ at time $t$''.  These markets will not only mediate
between the tag wearing individuals and tag reader operators, but also, and more
importantly, among reader operators, data mining service providers,
profile dealers and data customers after the information has been harvested.

The main and currently most influential industry standards for tags,
readers and number schemes are set by the industry consortium EPCglobal.  
Conforming EPC tags transmit an
\Em{Electronic Product Code (EPC)}, a data structure that has a globally
unique value on each tag (see Appendix \ref{app_epc_tags}).  
Where applicable, we use the terminology established by EPCglobal, our arguments, however, 
do not depend in general on any details in the standards. 
The central assumptions we use are made explicit in the following.

EPC tags for the mass market are small, short-ranged, and extremely
cheap RFID devices.\footnote{Reading ranges depend on frequency, antenna,
surroundings and energy.  Vendors are aiming at costs of less than
0.05\$ per tag for the mass market.} In particular, a tag cannot keep
track of who is reading it and when it is read, or selectively deny
reading. Standards for more capable tags have been proposed that can
selectively block read-outs or even communicate with trusted readers
only over encrypted radio links.  

Though the EPC Class 1 chip
specification includes a kill password to deactivate a tag permanently, this would 
prevent any after sale application, and the usability of these or comparable features
 is still unclear (see Section \ref{tags-are-simple}). We claim that collection and
pervasive distribution of tag data is essential to the usefulness of
RFID technology. Thus our arguments about the emergence of trace markets would still 
remain basically valid even in the presence of powerful and usable security
enhancements.

%According to EPCglobal, tags carrying unique EPCs will be attached to
%almost every object in everyday life. They can even be firmly
%integrated into clothing or other items, and could be nearly invisible
%to the human eye.
%
The envisioned ubiquity of tags, together with corresponding new (as well as existing) databases and
sophisticated data mining techniques, promises an enormous amount of
information on the movements of goods and individuals, 
and on correlations between entities and objects. 

However, very little work has been done on getting accurate or even
plausible estimates on the value of this information, the cost of
harvesting it, or on finding a model that describes what will happen
if this scenario becomes reality.  To confuse the situation further,
there are statements from industry leaders to the effect that interest
in this information is negligible,\footnote{See the statement of
the METRO Group during the consultation process by the EU Art. 29 Data
Protection Working Party:
\url{http://ec.europa.eu/justice_home/fsj/privacy/workinggroup/consultations/rfid_en.htm}}
despite efforts to create test beds that facilitate access to
it.\footnote{METRO Future Store, \url{http://www.future-store.org/}}

On the other hand, understanding the economics of EPC traces is crucial for
many reasons.  Policy-makers need to address the question whether
there is need for regulation.  Market players want to know the
ground on which they compete.  Finally, in order to estimate
the impact of Ubiquitous Computing technology on the privacy of
individuals, it is necessary to understand the motivations and incentives of
the parties that operate it.

The goal of our paper is to give a first description of EPC trace
markets, identify some of their most obvious characteristics, and lay
the ground for future research.

\XXX{check next paragraph before submission whether it's still valid
and all the pointers are good!} 

The paper is structured as follows: In Section
\ref{trace_markets} we give a brief overview of our terminology, the
market players, and the structure of goods.  In Sections \ref{demand} and
\ref{supply} we informally assess the demand and the supply side,
respectively.  In Section \ref{web_bugs} we compare the emerging 
EPC trace markets to the markets for online traces that exist today, 
drawing some conclusions on differences in interpreting information and on market risk.  In Section \ref{research_problems}, we
present further economic research topics that need to be addressed in order to develop a
deeper understanding of EPC trace markets.

%%%%%%%%%%%%%%%%%%%%%%%%%%%%%%%%%%%%%%%%%%%%%%%%%%%%%%%%%%%%%%%%%%%%%%%%%%%%%%%%%%%%%%%%%%%%%%%%%%%%
\section{Trace Markets}
\label{trace_markets}

The raw product of EPC trace markets is a database of entries of the form $(i, l,
t)$: $i$ is an identifier (in our case: an EPC); $t$ is a time
value (relative or absolute); and $l$ is a location
identifier (a GPS coordinate tuple, site of the supermarket, a
reader number, IP address, or any useful and
consistent description of the reader's context).
The original source of traces is the direct tag-to-reader interaction,
where a reader supplies power to all tags in range and collects their
EPCs. A reader could be a fixed installation, a hand-held device, or
it could be integrated into a mobile phone or PDA.

\begin{figure}[tbp]
\begin{center}
\caption{Players in EPC Trace Markets.}
\label{fig:model}
\end{center}
\end{figure}

We distinguish three main categories of market players (see Figure
\ref{fig:model}):

\begin{itemize}

\item The \Em{Individuals (Data Subjects)}. Entities that possess tags and generate traces. This category includes private citizens and employees of companies.

\item The \Em{Data Consumers}. The end users of trace data who require market profiles or
information on individuals for various reasons.  (See Section
\ref{demand}.)

\item The \Em{Data Suppliers}. Suppliers of trace data may be collectors who
operate tag readers and feed databases, or aggregators who provide
services by trading, pooling, and data mining such databases.  (See
Section \ref{supply}.)

\end{itemize}

The relationships and interactions between these categories are
complex for various reasons.

{\bf Individuals and Data Suppliers.} Information in this relationship can be
collected legally or illegally.  There are economic models in which
individuals face a trade-off, providing personal information
(disutility) in exchange for customized goods and services (utility).
Calibrating the involved utility functions is an ambitious goal.

{\bf Data Consumers and Individuals.} Being better informed about
the market, the data consumers are able to offer better products and
services for the individuals.  The information available from trace databases can greatly facilitate price discrimination. The reliability of data and the appropriate degree of price discrimination based on this is to be studied.

{\bf Data Suppliers and Data Consumers.} The players in this relationship no longer consider the EPC traces to be personal information. They rather regard them as tradeable information goods that can be sold to data consumers, who may then use the data to improve their own competitiveness in another, independent market. %This relationship is about information asymmetry and profit maximization that will require the implementation of design mechanisms. 

The authors of \cite{berendt:2001}\cite{shostack:2003}\cite{acquisti:2005} address the first two relationships. In this paper we focus on the
relationship between data suppliers and data consumers, whose interests 
disregard those of individuals.

\FUTUREWORK{fis@all: There is a distinction to be made between {\em
personalized data\/} that has been associated with an identity
implicitly or by inference (not by direct lookup), and {\em personal
data\/} that is explicitly associated with an identity.  If anywhere,
this should probably go here, but I don't think it's relevant to the
current structure of the paper.}

%%%%%%%%%%%%%%%%%%%%%%%%%%%%%%%%%%%%%%%%%%%%%%%%%%%%%%%%%%%%%%%%%%%%%%%%%%%%%%%%%%%%%%%%%%%%%%%%%%%%
\section{The Demand Side}
\label{demand}

There are indications that RFID traces in general and EPC traces in particular will prove to be instrumental to many parties. For example, a collection of possible uses is offered in an IBM patent application from as early as 2001 \cite{IBM:2001} (pointed out later by \cite{Albrecht:2005}):

\begin{quotation}
\noindent
{\em
In another embodiment, instead of determining the exact identity of the person, some characteristics such as demographic (e.g., age, race, sex, etc.) may be determined based on certain predetermined statistical information. For example, if items that are carried on the person are highly expensive name brands, e.g., Rolex watch, then the person may be classified in the upper-middle class income bracket. In another example, if the items that are carried on the person are "female" items typically associated with women, e.g., a purse, scarf, panty hose, then the gender can be determined as female.  [...]\/} (p.~2)

{\em When a person enters a retail store, a shopping mall, an airport, a train station, a train, or any location where a person can roam, a RFID-Tag scanner located therein scans all identifiable RFID-Tags carried on the person [...]\/} (p.~3)
\end{quotation}

This patent application gives anecdotal evidence that some experts
did foresee the potential usefulness of gathering quality information
through RFID traces.\footnote{Interestingly, IBM has also entered the
market for RFID privacy solutions later \cite{Karjoth:2005}.}  In this
section we try to give a more detailed view on the motivations of
potential trace consumers.  We consider companies, governments,
individuals, and research organizations.

%%%%%%%%%%%%%%%%%%%%%%%%%%%%%%
\subsection{Companies}

There are many reasons for the private sector to develop a substantial
demand for traces.  Even companies that gather their own supply of
traces will have reason to buy from or pool with other companies for
data completion, integration and refreshment.

{\bf Personalization.} There are empirically verified economic benefits
for companies to personalize their goods or services \cite{Pine:1993}. 
This has influenced industry best-practices \cite{Peppers:1999}. Especially personalization and
recommendation systems will be highly pertinent to in-shop or home RFID
applications (B2C) and will also increase demand for trace data in
E-Commerce. Some benefits of personalization are
(cf.~\cite{Schafer:1999}) the ability to turn casual browsers into
buyers, the potential of cross-sells by recommending matching items to
something already owned or selected by the customer, increased
customer loyalty and better customer relationship management.

Personalized insurances\footnote{An example today is ``Pay as you
drive``:
\url{http://www-1.ibm.com/services/ondemand/norwichunion.html}} will
stimulate a huge demand for traces by insurance companies to study a
person's whereabouts, movements and consumption habits \cite{Bohn:2004}.
Traces will also enhance the effectiveness of credit scoring
enormously by providing detailed insights into the subjects'
possessions and income. An indicator is today's market for corresponding 
information (see Appendix \ref{app_statusquo}).

{\bf Price Discrimination and Direct Marketing.} Price discrimination
has been identified as an important driver for privacy erosion on the
Internet \cite{Odlyzko:2003}. To maximize
profit a customer should ideally pay the maximum amount that is
acceptable to her.  In order to charge different customers different
prices for the same services or goods, data is needed to estimate
their willingness to pay. Data generated through personalization of shopping sites, click
tracing and other measures used on the Internet are conducive to such analysis \cite{Cranor:2003}.  

Traces will be a new source of relevant
information that also pertain to the physical world, answering interesting
product-related and customer-related questions:  Which products are likely to be bought together?
 What EPCs do people carry who do (do not) enter my store or buy my products?
 How do customers move in my store (all day)?
 What are the characteristics of my customers?  What do they already own?  What are their budgets?
 How do customers and non-customers differ?
 Which additional products should I offer to a customer?

If each customer in a shopping mall is recognized by the IT infrastructure as a walking list of the items she bought (nearly equivalent to a fully personalized customer), she is a much more convenient
target for product placement and direct marketing strategies.

{\bf Industrial Espionage.} Players in many industries will be tempted, if not actively interested, in the possibility of inspecting
a competitor's supply chain or simply lists of items or persons
that enter their buildings.  Also, less aggressive business
intelligence can make use of EPC traces as well: ``Which competitors
have made deals with which of my prospects? What kind of deals are these?''  Traces can make this information more accessible than it is today.

{\bf Intensified Competition for Information.} In a competitive environment where each
individual player can decide to increase her data pool in order to
increase her advantage over other players, an arms race in business
intelligence can emerge.
The following game can be used to model this effect.
Let the information held by company $i$ in the current round be $I_i$,
let $c, v$ be monotonously increasing functions describing the cost and
value of that information, respectively,
and let the utility $U_i$ of company $i$ be\footnote{A more accurate
utility function can no doubt be developed.  But as long as $U_i$
increases with $I_i$ and decreases with $I_j, j \not = i$, our
argument holds.}
\[
U_i = \frac{v(I_i)}{\avg_{j \not = i} v(I_j)} - c(I_i) .
\]
Then, if company $i$ increases its information in the next round to
$I^{*}_i = I_i + \delta_i$, company $j$ is worse off unless it
increases $I^{*}_j = I_j + \delta_j$ such that
\[
U^{*}_j = \frac{v(I^{*}_j)}{\avg_{k \not = j} v(I^{*}_k)}
         - c(I^{*}_j) \geq U_j .
\]

\XXX{seda@ben: ben can we talk about these formulas. it is not clear to me what I_k is, and there is no mention of what I_j is, or how the formula can be scaled to multiple companies???}

This holds assuming that all other players do not change their
information levels.  However, as now both companies $i$ and $j$ have increased
theirs, all other players have even greater pressure to do so as well in the
next round.

\XXX{Due to the lack of empirical data, $v, c$ are extremely hard to
configure, so this argument needs to remain speculative.  However,
...}

\XXX{fis@all: can we prove a suboptimal equilibrium here?  what does
incomplete information change in the strategies of the players?  who
knows anything about game theory?  (-:}

%%%%%%%%%%%%%%%%%%%%%%%%%%%%%%
\subsection{Governments}

For government agencies it will often be more convenient to accumulate raw or personalized traces from private companies, rather than to invest into additional reader infrastructures that cover sufficient area for permanent surveillance. 

Today, in the US,  government agencies often buy personal data from profile brokers like ChoicePoint.\footnote{References gathered by EPIC at \url{http://www.epic.org/privacy/choicepoint/}} We expect this trend to extend to the EPC traces at least until enough public readers are installed (e.g. for ticketing, traffic monitoring, billing and building security). We list some of the potential utilities of EPC traces for governments below.

{\bf Customs and Tax Collection.} Ownership of goods, their transfer and movement patterns are very interesting to custom services that could now track imported and exported goods. Likewise, tax collection for luxury items will be made easier by tracking items and their owners. Simply the threat of this possibility might reduce delicts and misdemeanor.

{\bf Law Enforcement.} The police will have a high interest in traces, as they will prove extraordinarily useful in forensics. They could be used to address questions like ''Where have these particular boots been sighted within the last month?''. Monitoring and remote surveillance of criminals or suspects will be made much easier.
Further, ideas like supporting civil disaster recovery plans through new technologies, e.g. in case of epidemics, may become possible once trace databases have become sufficiently large and accurate.

\FUTUREWORK{[fis] bruce schneier has given a talk at the almaden symposium
'privacy in data networks' (2003?) on the problems this will produce
for law enforcement: an abundance of false positives will hide more
information than can be recovered from the new data.  but this doesn't
fit so nicely in our train of thought, so let's just leave this
out. :-)}

{\bf Intelligence Agencies.} Even if they already have access to equivalent information that traces could deliver, traces could be used as confirming evidence to reduce uncertainties. Live traces could support other forms of surveillance, social (e.g. terrorist) networks could be analyzed more easily. Nevertheless, the challenge of false positives and false negatives will have to be tackled even more seriously with the increase of EPC traces.

\XXX{seda: \url{http://www.iht.com/articles/2006/05/14/news/intel.php};  fis@all: the article quotes schneier with his old rant.  nice.}

%%%%%%%%%%%%%%%%%%%%%%%%%%%%%%
\subsection{Individuals}
 
Individuals will also be interested in using trace databases.  This might be to quench natural human curiosity, or for more sinister activities like blackmailing or spying on neighbors, relatives or co-workers.  On the other hand, applications for child care as well as care for elderly persons make use of Ubiquitous Computing data, and are appreciated by their users for improving their quality of life considerably.
 
\subsection{Researchers}
 
Finally we foresee a substantial demand in raw data to support scientific research.  Examples include economics (e.g. improving research on trade), research in epidemics, migration and mobility research, and social sciences in general.

% one WEIS reviewer mentioned \cite{bloom}, and probably wanted us
% to reference it here.  but perhaps it's too far-fetched.

%%%%%%%%%%%%%%%%%%%%%%%%%%%%%%%%%%%%%%%%%%%%%%%%%%%%%%%%%%%%%%%%%%%%%%%%%%%%%%%%%%%%%%%%%%%%%%%%%%%%
\section{The Supply Side}
\label{supply}

The structure of the supply side of a market decides to what extent
anybody will engage in production.  In order to understand this, we look at the reader locations
where data is physically harvested, and the globally connected IT
infrastructure where it is exchanged and processed.  Then,
we assess potential obstacles that may perturb the trace supplier's job, namely technical and legal countermeasures.

%%%%%%%%%%%%%%%%%%%%%%%%%%%%%%%%%%%%%%%%%%%%%%%%%%%%%%%%%%%%%%%%%%%%%%%%%%%%%%%%%%%%%%%%%%%%%%%%%%%%
\subsection{Local EPC Traces}
 
In Ubiquitous Computing environments like smart buildings
RFID will be a fundamental technology to discriminate contexts. 
Applications like smart fridges, laundry machines and home medical
advisors could use RFID and the EPC Network to provide inhabitants with new services.  
Extensive reader infrastructures will be installed for these or other
primary purposes and may be used officially and with implicit consent for
trace harvesting. For example, these technologies can be employed inside a shop at the cashiers, scanning
and tracking the shop's EPCs (and those that the customers brought with
them). 

For service providers there lies plenty of opportunity for the secondary
use of these collected traces, which will often be already associated with
an individual (like in the case of smart homes) or easily
linkable (in the case of shop customers). As the market for traces develops, 
the reader operation is likely to be outsourced to specialized service providers. These will serve a greater set of customers in multiple settings and have access to larger volumes of data. If their contracts permit them to do so, these specialized service providers can also sell this data to formerly uninvolved third parties.

There also could be rogue readers operating despite the dissent of the
inhabitants of their environments.  Mobile or fixed installations
can be used to scan strategic areas within cities or buildings to gather
traces without consent or notice. More secretively, there is the
possibility to place passive sniffers\footnote{RFIDdump Project:
\url{http://rfiddump.org/}} near official reading
infrastructures to capture RFID communications and collect transmitted
EPCs.

Finally, creative marketing projects might come up with gadgets
like a portable music player that traces tags around its owner, producing more or less a complete profile of her and even of persons she knows.  
These mobile trace databases could be
uploaded to the vendor and traded for music or other services with
insignificant unit production cost.  Such business models could be
designed in the spirit of customer loyalty programs.

\subsection{Harvesting EPC Network Logs}
\label{ONS Insecurity} 

The EPCglobal network is designed to form a global information
retrieval network for objects carrying tags with EPCs. It follows the
{\em data-on-network\/} paradigm, i.e., information about objects is
not stored in the corresponding tags. Instead, these tags only contain
primary keys for data look-up from distributed
databases on the Internet, so called EPC Information Services (EPCIS).
The main function of the EPC Network is data exchange, first within
supply chains, then most probably as a backbone to an {\em Internet of Things\/} and to Ubiquitous Computing
environments. 

The EPC Network (like every solution for EPC Information retrieval via the Internet) 
will constitute an excellent trace trading system and, possibly even to a larger extent than local RFID interaction, a global trace-collection mechanism. Every EPCIS can itself be a trace collection point, simply by log file analysis. Instead of the physical location an EPCIS collects the source IP
of the query, and probably further identifiers and authentication tokens from the
information protocol layer. Third parties cannot sniff this trace-containing traffic
if it is encrypted.

But the look-up service {\em Object Naming Service (ONS)\/} is a weak
spot in the proposed design \cite{Fabian:2005}. ONS uses the structure of an
EPC, which is transformed into a domain name, for query delegation and
in functionality is akin to the Internet's Domain Name System (DNS). Every ONS server
in the resolution chain is a potential trace collection
point.\footnote{The ONS specification as of today does not use the
serial number for delegation, but leaves it as option for the future.
In its current form, identifying and tracing rare item classes or
clusters of items will be possible.}
ONS is a clear text protocol and is about to inherit all the well-known security weaknesses of DNS. In the worst (or best, respectively)
case, every IP node in the ONS resolution chain, every router
or network analysis device in the path will be able to collect partial traces.
%\XXX{mb: Koennen wir das DNSSEC rauswerfen? Das taucht immer nur
%als hypothetische Loesung auf, aber niemand tuts, weils sehr komplex ist }
%
%DNSSEC could be used
%to mitigate some of these risks, but is until today not globally
%deployed. To hope for the widespread use of DNSSEC for ONS sometime in the
%future might turn out as wishful thinking. But, even then the encryption of communication in DNSSEC remains a problem. 

We anticipate that EPC information exchange over the Internet, 
and especially the EPC Network, will make traces even more widely available, 
in addition to those generated by local and direct skimming via RFID readers.  
Collection and data
aggregation systems like intrusion detection systems that are in
place today could be modified to harvest EPC trace data from the
Internet.  Combining these into large, global aggregation services
that exist for, for example, distributed intrusion detection\footnote{DShield:
\url{http://www.dshield.org}} today, or including them in search
engines like {\em Google} or RFID-pendants of services like {\em
Where's George}\footnote{\url{http://www.wheresgeorge.com/}}
might let lookups for
specific EPC clusters succeed with high probability.

\subsection{Restricting Supply}

%In principle, the society can pick from two strategies to keep people from
%doing something deemed as undesirable: Make it impossible (or at least
%cumbersome and expensive) to do, or make it illegal. 
In the following sections, we look at the strategies for restricting the unwanted use of EPC traces and study their possible repercussions for the trace trade markets.

\subsubsection{RFID Protection: Restricting the Information Flow}
\label{tags-are-simple}

If tags are ubiquitous, a lot of privacy issues arise 
\cite{Albrecht:2005}\cite{Garfinkel:2005a}\cite{Stajano:2005}.
Several security enhancements to control the flow of information have
been proposed to maintain a certain level of privacy for
individuals \cite{Garfinkel:2005a}\cite{Juels:2006}.\footnote{A
dedicated Web site for RFID security and privacy research is
maintained by G. Avoine: \url{http://lasecwww.epfl.ch/~gavoine/rfid/}}

In order to estimate the density of traces available for
harvesting and trading, one needs to understand these security
enhancements.  In this section, we give a very brief overview of some existing
data protection methods and identify problems that may
negatively affect their efficiency in practice.

Tag readers are publicly available
and will become cheap. Hence, anyone will be able to read all the tags located within a few meters
distance and with minimal effort
maintain traces of the area around her readers.  Consequently, not
only could operators of legitimate readers re-use the data for
trading, but there could be reader operators that collect traces
solely for the trace markets. Various enhancements to block or control read-outs have been proposed in order to confine the set of readers to those deemed legitimate by the individual wearing the tags.  Although promising, these solutions face a number of obstacles.

{\bf Selective Deactivation or Blocking.} This approach is
inconvenient and (without strong authentication) highly problematic. 
If a chip
needs to be disabled every time a hostile reader might be nearby, and
re-activated every time a friendly reader is expected to read it,
things will frequently go wrong.
The user will forget to disable tags, or hostile readers will be
installed close to friendly readers, and the user will have to choose
between not profiting from the tag infrastructure and being exposed to
a hostile read-out.

{\bf Symmetric Cryptography.} Authenticating readers (possibly also the tags) 
and communicating over an encrypted channel
can make blocking out untrusted readers both more convenient and more
reliable.  Symmetric cryptography can be used for both authenticity
and encryption.

Since the storage capacities of a tag are extremely limited,
only a small number of symmetric keys can be stored on a tag.  All
readers potentially communicating with a tag need to possess one of
the keys stored on the tag.  This establishes \Em{trust clusters}
among large numbers of readers and huge numbers of tags.  Each message
can only be shown to originate from a certain cluster, not from a
certain device.  Consequently, any adversarial trace collector with
physical access to any tag or reader from a given trust cluster can
retrieve the symmetric key for that cluster, pretend to be a trusted
reader, and harvest the protected tags.

{\bf Asymmetric Cryptography.} Whereas symmetric authentication
and encryption in applications with many communication partners
quickly exhaust the key management capabilities of the 
parties involved, asymmetric cryptography exceeds the computing resources of
most RFID tags. Especially tags that have no batteries and need to be
small and cheap enough to tag small and cheap goods are not suitable for the implementation of such mechanisms.  Asymmetric cryptography is at best
considered an option for tags with higher capacity used in production and logistics.

{\bf Trust.} A more exotic option is to broadcast privacy policies from a PDA
\cite{Floerkemeier:2005}.  This PDA would ask all readers in scanning
distance to ignore all sensitive EPCs, and law enforcement will
provide the incentives for reader operators to obey the policies that
are expressed. This approach is likely to fail in the trace application scenario for at least
two reasons: First, economic incentives to get to the data are
strong, as this paper demonstrates; second, with P3P, an existing
system based on this idea has proven too inconvenient for the user to
be widely deployed \cite{hochheiser}.  So in practice, neither opt-in
nor opt-out are easy options. 

Writing privacy policies is complex and cumbersome.  
More importantly though, content providers on the Web have the power to make it inconvenient for users to choose a restrictive privacy
policy.  Usually they can do this by configuring unnegotiable policies, leaving it to the user to either defer the use of a site, or to relax her own preferences.  In the RFID future
we envision, those players planning EPC-based services will be
able to reiterate this practice, but then not only Web
browsing behavior, but rather the activities of the user in the physical world will be affected.

%{\bf Link between Reader and Company IT Infrastructure.}
%hackers?  bribing?  not the weakest link, though.

{\bf Link between Company and EPC Network.} The security of current EPC Network and ONS
drafts is, to say the least, controversial (see Section \ref{ONS Insecurity} above).  So
even if the link between tag and reader would be perfectly secure, the
information flowing through the global EPC infrastructure (of which readers
are only the leaves) provides a potentially even richer field
for trace collectors and brings new challenges to IT security.

\subsubsection{Privacy Laws and EPC Traces}
\label{legal}

In some countries there are laws that restrict the ways in which governments, corporations, and citizens are allowed to collect, store, distribute and make use of personal information. However, the raw traces under scrutiny here are not personalized {\em per se}. They can only provide privacy-relevant information by use of data mining techniques and the consultation of further data sources (such as a data warehouse that may contain a mapping from EPCs to customer IDs). Each single act of trace harvesting that is noticed and brought before a court will likely not have caused any damage on its own, as the data is only valuable in large amounts. Therefore, it is unclear whether trace collection is covered by privacy laws at all.

{\bf The Question of Enforceability.}
Suppose the problem becomes apparent to the public and to policy makers in the future and a law prohibiting trace harvesting and trading is passed (despite the economic advantages of collecting such information). If the incentive structure is in favor of trace trade, an illegal market will emerge in substitution of the legal markets of traces, just like the markets for illegal drugs, botnets, or credit card numbers \FUTUREWORK{give references to literature here?} are perfectly healthy and operational despite the laws against them. Laws can affect the cost of supply (i.e., the price of the trace harvesting and trading infrastructure) and the product quality (i.e., the trace density), but whether the impact on trace markets in particular will be high enough to stop trade is uncertain.

%There have been laws against spam and telemarketing in several countries for years now, but the industry is still prospering, because it is impossible to prove who out of a large number of legal bodies made a certain phone call or sent a certain e-mail. 
%
%Networks of globally operating companies have emerged that are exceedingly cumbersome to investigate and blame for any particular action or event. In contrast to spam activity, the effects of EPC trace collectors are harder to notice for the individual.

{\bf De-Anonymization.}
Privacy laws usually introduce some notion of {\em how personalized\/} information is. For example, customer profiles that are aggregated to one tuple for every five households may be considered anonymized, or a certain volume of trace information may be considered harmless. However, in practice even a single aggregated data source often contains enough information for a good data analyst to de-anonymize it to a large extent.\footnote{A much quoted figure is that $87\%$ of all Americans are uniquely identifiable from ZIP code, birth date, and sex, so aggregation needs to obfuscate this information \cite{Sweeney:2002}.}  A data pool of traces and other corpora that is too enriched to be legally owned by one company can simply be split up between networks of companies, and the knowledge can be combined in a way that is certain to go unnoticed until aggregate data emerges in a completely different context.

{\bf From Personal Data to Intellectual Property.}
There may even be laws such as intellectual property rights antagonizing privacy regulations. Once a market research agency has aggregated the collected trace data and thereby turned it into intellectual property, it is not only accessible to paying third parties, but also harder to obtain by the affected individuals. 

{\bf Explicit Consent where there is no Contract?}
Explicit consent to collection of personal data can of course not be given if there is not even a contractual relation between the profiled individual and an independent trace collector operating in public spaces, or in private spaces not owned by him. On the other hand, users have to accept the terms imposed by the owner of a space they enter, for example video surveillance in shopping malls. By analogy, it seems likely that the owner of a shopping mall may legally decree that all goods sold on its premises are equipped with RFID tags, and that tracking is performed.

\FUTUREWORK{Amerik. provider verkaufen handy profile (mb):
http://www.suntimes.com/output/news/cst-nws-privacy05.html}

%%%%%%%%%%%%%%%%%%%%%%%%%%%%%%%%%%%%%%%%%%%%%%%%%%%%%%%%%%%%%%%%%%%%%%%%%%%%%%%%%%%%%%%%%%%%%%%%%%%%
\section{EPC Traces vs.\ Web Traces}
\label{web_bugs}

We will now take a closer look at an already existing information market that might prove insightful to EPC traces markets, although it seems to have been exposed to only sparse economic research: The market of Web Analytics.\footnote{This market is changing at the moment, because a big player has entered the stage: Google now provides analysis of
click traces for free (Google Analytics \url{http://www.google.com/analytics}).  Such services are possible for EPC traces as well.}  Visitors of the World Wide Web produce traces in server logs that are often subject to complex business models.
One special tracing method is the so called \Em{Web Bugs}\footnote{EFF Definition: \url{http://www.eff.org/Privacy/Marketing/web_bug.html}},
nearly invisible linked images that allow tracking of users across different Web sites or via electronic mail.  Among other things, they are used to verify how many
users viewed an ad or read an e-mail. Consequently, despite privacy concerns, they constitute a popular basis for advertisement pricing and market research.

Web traces are similar to EPC traces, the EPC corresponds to a
client IP (and often an additional session ID), the location
information obtaining the form of a URL being accessed.
We argue that EPC trace databases will be similarly successful, if not
more so, as they mirror movements in the physical world.  However, they differ in important details.

{\bf Data.} EPC traces per se do not show a sequence of volitional
events (such that, e.g., a preference for two products must be
identified from their relative sequential positions) but a
simultaneity of relations to things (e.g., carrying around two
products together on mornings, but never together in the
afternoons).

{\bf Aggregation Technology.} EPC traces bring about new algorithmic challenges:
Needed are algorithms on spatio-temporal data of high resolution;
data integration to obtain a semantic and thus actionable results;
and algorithms that work on data with a rich relational structure.

Such algorithms have been developed over the past decades. 
They align different granularities of spatial data (e.g., geographical
coordinates and street networks), they can integrate the semantics of
space and also of action models (e.g., a person's home and work
location, different kinds of trips), for a current example see 
\cite{AAAI04-paper}. Advances in
\mbox{(multi-)relational} data mining \cite{multi-relational} are also substantial.  Thus, if
the supplier of EPC traces offers analysis options, the costs for
additional intelligence are low.
% If she only offers data, the cost
%of analysis does not enter the supplier's cost calculation at all.

{\bf The Passive Role of the User.} EPC traces arise from ubiquitous
contexts and not from limited interaction with software via an interface
that can be configured by the user.  
%As explained in Section \ref{tags-are-simple}, we assume that tags are passive and
%broadcast in response to requests from all appropriate readers. 
The consequences are the occurrence of more traces, and in principle traces that have higher density
than the corresponding traces collected on the Web. On the other hand, the interpretation of what an
ID-time-location triple means may become even harder than with
click traces.
%, where in general it can not easily be determined why something was
%chosen.
In the Web trace business, errors are sometimes intentionally fed into the databases by privacy-aware users. Another aspect of the passive role of the user in the collection of EPC traces is that she cannot effectively lie about the data that is generated.  

{\bf The Problem of Coverage.} The spatial entities
that are being tracked are different, and so is the locus of
control. 
%In essence, click tracing starts from {\em locations} that
%are comprised of {\em objects} that customers
%potentially interested in data on these objects own(locations are Web
%sites, objects are products and services offered on these sites, and
%customers are the site owners).  
EPC tracing also starts from
locations, but these are usually public and exist independently of the objects that are being
tracked.  In the absence of universal tracking coverage, the collectors of traces cannot guarantee the collection of data that might be of interest to a given customer.  

This will be mitigated by the following circumstances:
First, the trace collector may be the owner of the space under
surveillance.  In this case, the situation is very similar to the Web, the market risk is low, and good coverage is affordable. 
Second, the trace collector may have entered into a contract to collect traces at specific locations. 
The risk that no interesting data are collected must be borne by either or both of the contract
partners, and appropriate pricing models will need to be developed. 
Third, these bootstrapping problems do not exist if companies already gather
data for purposes other than trading, yielding tradeable traces as a
positive externality. 

%Companies or governmental agencies may start out by gathering their own supply of traces. Eventually, they will have incentives to pool this information with other companies or to demand more diverse data from intermediaries. The demand for more diverse, complete and fresh data may then be a driving force of this market. 

%In another scenario, early locations may be limited to expected loci of interest (like main squares, shopping centers, airports). In both cases, a wider market may first emerge once the demand for traces and personalized data is better defined and a need for location-diversity is articulated by the customers.

Summing up, the collection and analysis of EPC traces seem to imply only manageable additional business risks compared to Web tracing, but will encounter much higher demand, as there are more customers interested in tracing the physical than the virtual world.

\section{Economic Research Problems}
\label{research_problems}
 
In the following we discuss some issues that may arise in the EPC markets between the data consumers and the data suppliers, and identify future research questions in modelling this relationship.

{\bf Web Trace Markets.} What is the state of Web trace markets today, how much money is made? Similarities and differences to EPC trace markets need to be studied further. 

{\bf Initialization of EPC Trace Markets.} We introduced some initialization scenarios in Section \ref{web_bugs}. During the initial phases of the market it may not be clear how suppliers of EPC traces can best select the locations of their readers and find the appropriate customers for their trace collections or aggregated data. 
%Hence, the market may first have to be initialized through adequate business models and design mechanisms.

{\bf Information Asymmetry.} Contracts between trace collectors and trace consumers will include information on target EPCs and locations. Customers of data will reveal their specific interest in the process, while the supplier won't give data away indiscriminately.  This could cause contracting problems and may result in information asymmetries that may lead to market failures. The implementation of design mechanisms to avoid such a market failure is a topic of future research.

{\bf Quality Assurance.} The market of EPC traces will be vulnerable to malicious actors who sell fake traces. Competitive partners may be interested in injecting false data in order to effect the analyses of their competitors. This may lead actors utilizing the traces to false analyses and decisions, and may have economic repercussions.

%When fake data is available and detected, quality and reliability of data may become an important aspect of the market. Buyers of trace collections or aggregated data may demand some sort of data authentication. This may establish a hierarchy of trusted and untrusted data providers. Security mechanisms and reputation models may become an integral part of guaranteeing quality and reliability in the market, but will effect the prices of trace collections and resulting data. Understanding these cost factors and their effect on the markets remains an open research topic.

%\input{p06_model}

%%%%%%%%%%%%%%%%%%%%%%%%%%%%%%%%%%%%%%%%%%%%%%%%%%%%%%%%%%%%%%%%%%%%%%%%%%%%%%%%%%%%%%%%%%%%%%%%%%%%
\section{Conclusions}

Under the assumptions that networked RFID technology  and usage of the Electronic Product Code in general become ubiquitous, we have argued that there is a strong incentive to aggregate and subsequently trade traces of EPCs.  

We have assessed the nature of the business models involved and investigated potential buyers of trace-based products. A closer look at the supply side brought us to the preliminary conclusion that despite potential technical and legal barriers trace markets will emerge. 

Comparing the trace market to Web Bugs has given indicators that business risk for data providers and market initialization will be manageable. Further, we have sketched some topics for further economic research concerning the relation of data consumers and providers.

Evaluating our results in the broader context of society, we come to the conclusion that ultimately a balance between two extreme states needs to be found: One extreme would be a state in which only secure RFID technology is introduced that doesn't produce much information on individuals or goods. Given the current technological developments this state is impractical and economically ineffectual. In the other extreme state security is not an issue, information flows perfectly free (and hopefully for the benefit of both individuals and the corporate world), but privacy is drastically eroded by the emerging information markets.

Crucial questions are: Where do we, as individuals, want to be between these two possible states in our daily lives, potentially {\em despite} the market forces? And will scalable and usable security and privacy enhancements to RFID and EPC systems be implemented to get us there?

\newpage

\section{Acknowledgments}

We would like to thank Steffan Baron, Bettina Berendt, Holger Gerhardt, Michael Klafft,
Sushmita Swaminathan, and the anonymous reviewers of the WEIS'06
workshop for valuable comments on earlier versions of this paper.

\bibliography{bib}
\newpage
\appendix

\section{Appendix: Profiles as Products, Today}
\label{app_statusquo}

RFID and EPC traces will increase the granularity of profiling individuals, creating new paths for data mining to extract high level information.  For some data there already exists a market today. 

The following is a partial price list (May, 2006) cited from the SWIPE project (\url{http://www.turbulence.org/Works/swipe/}). The prices are collected from commercial warehouses and profile brokers (Accurint, Aristotle, ChoicePoint, ChoiceTrust, DocuSearch, Experian, KnowX, Merlin Data, and Pallorium). All prices in \$.

\begin{tabular}{|l|l|}
\hline
General & \\
\hline
Address & 0.50 \\
Zip Code & 0.50 \\
Past Addresses & 9.95 \\
Date of Birth & 2.00 \\
Marriage & 7.95 \\
Divorce & 7.95 \\
Education & 12.00 \\
Employment & 13.00 \\
Published Phone \# & 0.25 \\
Unpub. Phone \# & 17.50 \\
Cellular Phone \# & 10.00 \\
Past Phone \#s & 0.50 \\
Relatives & 3.00 \\
Neighbors & 0.25 \\
Registered URL/Domain Name & 0.25 \\
Soc. Security \# & 8.00 \\
\hline
\end{tabular}
\begin{tabular}{|l|l|}
\hline
Financial & \\
\hline
Credit & 9.00 \\
Real Estate & 1.50 \\
Bankruptcy & 26.50 \\
Worker's Compensation & 18.00 \\
Assets & 6.95 \\
Assets Seized & 2.95 \\
Shareholder & 1.50 \\
Executive Affiliation & 0.50 \\
Own Aircraft & 1.50 \\
Own Boat & 1.50 \\
Own Vehicle & 0.75 \\
Own Business & 9.95 \\
\hline
\end{tabular}

\begin{tabular}{|l|l|}
\hline
License Info & \\
\hline
Driver Lic. Info & 3.00 \\
Motor Veh. Reg. & 3.00 \\
List of Vehicles & 0.70 \\
Accident Reg. & 1.00 \\
Aircraft Lic. & 1.50 \\
Drug Enforcement Admin. (DEA) Lic. & 0.25 \\
Hunt \& Fish Lic. & 0.25 \\
Professional Lic. & 0.75 \\
Industry Accreditation & 16.00 \\
Merchant Vessel & 0.25 \\
Concealed Weapon & 0.25 \\
Firearms Lic. & 0.25 \\
\hline
\end{tabular}
\begin{tabular}{|l|l|}
\hline
Legal & \\
\hline
Lawsuits & 2.95 \\
Felony & 16.00 \\
Misdemeanor & 9.00 \\
Sex Offender & 13.00 \\
\hline
\end{tabular}
%\begin{tabular}{|l|l|}
%\hline
%Political \& Military & \\
%\hline
%Voter Reg. & 0.25 \\
%Military Record & 35.00 \\
%\hline
%\end{tabular}

\section{Appendix: Electronic Product Code (EPC)}
\label{app_epc_tags}

The most prevalent kind of EPC (SGTIN-96) has a length of 96 bit and
corresponds to a SGTIN (Serialized Global Trade Identification
Number). The main structure of this EPC is as follows
\cite{EPCglobal:2006} (see Figure \ref{EPC}):

\begin{figure}[h]
\begin{center}
\caption{Electronic Product Code (SGTIN-96 EPC)}
\label{EPC}
\end{center}
\end{figure}

The Header defines the kind of number that is encoded (here
SGTIN-96). Filter Value is a rough object classification (e.g., retail
item). Partition determines the exact boundary between the two following
values.  Company Prefix (former EPC Manager) determines which company
issued this EPC (usually the manufacturer of the corresponding item).
The Item Reference (Object Class) determines the exact category that
the tagged object belongs to, and the Serial Number identifies a
particular item within the same object class.  

Note that this serial number enriches the information carried by an EPC tag significantly
compared to a bar code, making distinction between individual items of the same kind possible.

\end{document}